\documentclass[aps,jmp,amsmath,amssymb,preprint]{revtex4-1}
\usepackage{graphicx}
\usepackage{bm}
\usepackage{mathptmx}
\usepackage{epstopdf}
\usepackage{booktabs}
\usepackage{multirow}
\usepackage{hhline}
\usepackage{subfigure}
\usepackage[usenames,dvipsnames]{color}

\begin{document}
\title{Cooperation and competition between magnetism and chemisorption}
\author{Satadeep Bhattacharjee$^{1}$}
\email{s.bhattacharjee@ikst.res.in}
 \author{Seung-Cheol Lee$^{1,2}$}
 \email{seung-cheol.lee@ikst.res.in}
\affiliation{$^{1}$Indo-Korea Science and Technology Center (IKST), Bangalore, India \\
  $^{2}$Electronic Materials Research Center, Korea Institute of Science $\&$ Tech., Korea}  
\begin{abstract}Chemisorption on ferromagnetic and non-magnetic surfaces is discussed within the Newns-Anderson-Grimley model along with the Stoner model of ferromagnetism.
In the case of ferromagnetic surfaces, the adsorption energy is formulated in terms of the change in surface magnetic moments. Using such a formulation, we address the issue of how an adsorbate's binding strength depends on the magnetic moments of the surface and how the adsorption process reduces/enhances the magnetic moments of the surface. Our results indicates a possible adsorption energy scaling relationship in terms surface magnetic moments. In the case of non-magnetic surfaces, we formulate a modified stoner criterion and discuss the condition for the appearance of magnetism due to chemisorption on an otherwise non-magnetic surface.
\end{abstract}
\maketitle
\section{Introduction}
Interfaces of the gas molecules and the metal surfaces are the basis of heterogeneous catalysis and molecular spintronics. Conventional catalysts are non-magnetic heavy metals ~\cite{conv1,conv2,conv3}, and are not interesting platform for discussing the magnetic properties. The focus has, however, turned in recent years towards simple 3d-transition metals to be used as catalysts~\cite{gercia}. In most of the cases 3d transition metals are used as alloying elements in either binary or ternary alloys-catalysts~\cite{S1,S2}. These materials offer additional degrees of freedom such as internal strain and magnetic moment when alloyed with traditional heavy metals.
Further more if the gas molecule is paramagnetic (such as oxygen), the spin dependent interaction between the metal and molecule becomes important ~\cite{gercia}.
Therefore effect of magnetism on surface reactivity is gradually becoming focus of  many recent studies~\cite{f1,f2,f3}. Also, changes in the magnetic properties of the metals induced by the adsorption of the gas molecules promise various new avenues particularly in the field of spintronics~\cite{av1,av2,av3,av4,av5}.

The effect of surface magnetism on chemisorption and the impact of chemisorption on surface magnetism are both interesting subjects and demand thorough theoretical discussions. In the literature, we find the mention of both cooperation and rivalry between the magnetism and chemisorption: (1) On the one hand, it has been seen that the saturation magnetic moment of the Ni surfaces decrease due to chemisorption of gases~\cite{Ni-gas1,Ni-gas2}. (2) On the other hand, some recent works suggest that the chemisorption of molecules helps the non-magnetic metals to overcome the Stoner criterion and make them ferromagnetic at room temperature~\cite{beating}. Similarly chemisorption induced ferromagnetism  is reported in Au and Pt nanoparticles~\cite{mujika,Sakamoto} 
%\par
%From the theoretical side, the chemisorption can be described by the  Newns-Anderson-Grimely model ~\cite{Newns,Anderson,Grimley} while the metallic ferromagnetism via Stoner model~\cite{Stoner}. Within the Stoner model, the off-set of magnetism depends on the density of states (DOS) at Fermi energy while the covalent metal-adsorbate binding is described by Newns-Anderson-Grimeley approach in terms of a coupling between the adsorbate states with the metal states.
\par
It is well-known that chemisorption of the molecules effects the density of states  of the metals near the Fermi-energy. For example, it was proposed that the decrease of the magnetic moment on the Ni surfaces due to chemisorption of gases are result of the shift of the DOS towards the foot of the d-bands~\cite{Ni-gas1,Ni-gas2}. The degrees of adhesion of an adsorbate on a metal surface is understood in terms of \textit{adsorption energy}. While it is well established that the electronic structure of the metal surface changes due to chemisorption, there is no simple mathematical relationship between the adsorption energy and the change in magnetic moment in the literature.

In this manuscript, we formulate a mathematical relationship between the adsorption energy and the change in the surface magnetic moment (which occurs due to chemisorption) of a ferromagnetic metal surface. Keeping such formulation as a basis, we try to understand the experimental findings as mentioned in the point (1), above. 

To understand the experimental situations as mentioned in point (2), we move a little further, we derive a modified Stoner criterion resulting from chemisorption on non-magnetic metal surfaces. Such a formulation will enable one to understand how the standard Stoner criterion can be violated, as shown, for example, in Ref.[\citenum{beating}].

We organize our paper in the following way: In the section 2, we first make a general formulation of adsorption energy in terms of the chemisorption induced change in density of states and magnetic moment. We, then, describe the metal surface in terms of Stoner Hamiltonian with Bloch states as basis while the adsorbate is described in terms of a localized states and the coupling between the localized and extended states were introduced as in Newns-Anderson-Grimley model. Using such an approach we finally derive an expression for the adsorption energy which is dependent on the change in the magnetic moment of the metal surface due to the adsorption. The interdependence of the magnetic properties and chemisorption are then evaluated by comparing two phase diagrams obtained from the same model: one is a magnetic phase diagram which describes how magnetic moments change due to chemisorption, the other is a chemisorption phase diagram describing how chemisorption energy changes with respect to change in surface magnetic moments. Both phase diagrams are obtained by varying the same sets of parameters with values within the same range. In the section 3,  we deduce a modified Stoner criterion and  analyze how chemisorption may help to overcome the barrier laid by the standard Stoner criterion. The Stoner criterion is the most important result of the Stoner theory of band ferromagnetism which states that the ferromagnetism in the metals originates from a competition between the band energy and the exchange energy. The spontaneous onset of magnetism within this model is given through, $ID(E_F)\geq 1$. Here I is the Stoner parameter and is related to the exchange splitting of the bands, and $D(E_F)$ is the density of the states at Fermi energy.
In the present work, we focus on the cases where the metal surfaces do not meet the Stoner criterion without any disruption, but may become ferromagnetic due to chemisorption.
In the section 4, we made the conclusions.
\section{Chemisorption on ferromagnetic surfaces}
\subsection{General relation between the adsorption energy and change in surface moments}
Let $m_i$ is the magnetic moment per site of the ferromagnetic surface and $m_f$ is the magnetic moment after the adsorption happened. If $D_\sigma(E)$ and $\tilde{D}_\sigma(E)$ are the DOS of the metal surface before and after the adsorption ($\sigma$ is the spin index), then the change in magnetic moment ($\delta m=m_f-m_i$) after the chemisorption is given by,
\begin{equation}
\delta m=\int ^{E_F}_{-\infty} \left[\tilde{D}_\uparrow(E)- D_\uparrow(E) \right] dE-
\left[\tilde{D}_\downarrow(E)- D_\downarrow(E) \right]dE
\label{mom1}
\end{equation}
Where $E_F$ is the Fermi energy. If we use $\tilde{D}_\sigma(E)=D_{\sigma}(E)+\Delta D_\sigma(E)$ as DOS of the metal surface after the adsorption, then the change in magnetic moment can be written in terms of the  change in the density of states: $\Delta D_\sigma(E)=\tilde{D}_{\sigma}(E)-D_{\sigma}(E)$. 
\begin{equation}
\delta m=\int ^{E_F}_{-\infty}\left[ \Delta D_\uparrow(E)-\Delta D_\downarrow (E)\right]dE
\label{mom2}
\end{equation}
The adsorption energy of an adsorbate with renormalized energy level $\varepsilon_{a\sigma}$ is given by,
\begin{equation}
\begin{split}
\Delta E (\delta m)&=\sum_\sigma \int_{-\infty}^{E_F+\delta E_F}E\tilde{D}_\sigma(E)dE-\sum_\sigma\int_{-\infty}^{E_F}ED_{\sigma}(E)dE-\sum_\sigma n_{a\sigma}\varepsilon_{a\sigma}\\  
& = \sum_\sigma \int_{-\infty}^{E_F} E\Delta D_{\sigma}(E)dE+E_F\left[\delta m-2\int_{-\infty}^{E_F}\Delta D_\uparrow(E)dE\right]\\
&+\sum_\sigma n_{a\sigma}(E_F-\varepsilon_{a\sigma})
\end{split}
\label{EB1}
\end{equation}
$\delta E_F$ is the small variation of the Fermi energy upon adsorption, $n_a$ is the number of electrons in the adsorbate. To derive the Eq.\ref{EB1}, we have used the Eq.\ref{mom2}, and the charge neutrality condition as is given by
$$\sum_\sigma \int_{-\infty}^{E_F+\delta E_F}\tilde{D}_\sigma(E)dE-\sum_\sigma\int_{-\infty}^{E_F}D_{\sigma}(E)dE=\sum_\sigma n_{a\sigma}$$
The Eq.\ref{EB1}, gives the adsorption energy of a molecule on a ferromagnetic metal surface with  spin dependent density of states $\Delta D_\sigma(E)$. The first term in the equations refers to the change in the band energy due to the change in DOS, while second and third term are related to the change in the magnetic moment of the metal surface due the chemisorption. The last term in the Eq.\ref{EB1} refers the energy separation between the Fermi energy of the metal and the adsorbate energy level. It is clear that for a metal surface with positive Fermi energy, if the chemisorption enhances it magnetic moment ($\delta m>0$), this would lead a smaller(less negative) adsorption energy. While if the chemisorption reduces the magnetic moment  ($\delta m < 0$), the chemisorption energy will depend on the magnitude of the reduced moment.

For the non-magnetic surfaces,$\delta m=0$ and $2\int_{-\infty}^{E_F}\Delta D_\downarrow(E)dE=\int_{-\infty}^{E_F}\Delta D(E)dE$, therefore the Eq.\ref{EB1} reduces to the 
well-known \cite{AP}form,
\begin{equation}
\Delta E= \int_{-\infty}^{E_F} (E-E_F)\Delta D(E)dE+\sum_{\sigma}E_F(n_{a\sigma}-\varepsilon_{a\sigma})
\label{wellknown}
\end{equation}

%To illustrate how $\Delta E$ depends on $\delta m$ and vice versa we solve numerically the Eq.\ref{mom2} and Eq.\ref{EB1}. This requires us to know $\Delta D_\uparrow (E)$,~$\Delta D_\downarrow (E)$ and eventually $\Delta D(E)$.\\
%that would lead a relatively larger (more negative) adsorption energy. This predictions are in agreement with the experimental findings~\cite{Ni-gas}
\subsection{$\Delta E (\delta m)$ within Newns-Anderson-Grimley-Stoner framework}
To calculate numerically the adsorption energy for a ferromagnetic metal surface we need to know $\Delta D_\sigma(E)$ which we obtain within a framework which combines the Newns-Anderson model with Stoner model as follows.\\
The Hamiltonian for the metal surface can be written in terms of the Bloch states \cite{St1,St2}
\begin{equation}
\hat{H}_{M}=\sum_{k j \sigma }\left(\varepsilon_{k j}+I\langle n_{j-\sigma}\rangle\right)c^\dag_{kj\sigma}c_{k j \sigma}
=\sum_{k j \sigma } \varepsilon_{k,j,\sigma} c^\dag_{kj\sigma}c_{k j \sigma}
\label{stoner}
\end{equation}
where I is the Stoner parameter and j refers band index. The adsorbate Hamiltonian can be written as 
\begin{equation}
\hat{H}_{ad}=\sum_{\sigma}\varepsilon_{a\sigma}\hat{n}_{a\sigma}
\label{ads}
\end{equation}
Here $\varepsilon_{a\sigma}$ is the adsorbate energy level with occupation ${n}_{a\sigma}$.
The coupling between the two subsystems (adsorbate and the surface) is given by,
\begin{equation}
\hat{H}_{C}=\sum_{k,j,\sigma} V_{k,j,\sigma}c^{\dagger}_{k,j,\sigma}c_{a\sigma}+h.c
\label{coup}
\end{equation}

The unperturbed Hamiltonian of the system can be written as, $\hat H_0=\hat H_M+\hat H_{ad}$. While the total Hamiltonian of the system is given by $\hat H=H_0+H_C$. The change in density of states referred in the Eq.\ref{EB1} can be obtained as,
\begin{equation}
\Delta D_\sigma(E)=\Delta n_\sigma(E)+\delta(E-\varepsilon_{a\sigma})
\end{equation}
where $\Delta n_\sigma(E)
=-\frac{1}{\pi}\Im\left (G_\sigma(E)-G^0_\sigma(E)\right) =-\frac{1}{\pi}\Im\left [\frac{d}{dE}ln Det(1-VG^0_\sigma)\right]$ 
. $G_\sigma(E)$ and $G^0_\sigma (E)$ are  the retarded single electron Green's function of the coupled and decoupled metal adsorbate system. 

It can be shown that, for a particular spin, $\sigma$ the change in DOS can be written as~\cite{AP} (also refer supplemental material),
\begin{equation}
\Delta D_\sigma(E)=-\frac{1}{\pi} \Im\left [(1-\frac{d\Sigma(E)_\sigma}{dE})G_{\sigma,a}(E)\right]
\label{change}
\end{equation}
where $\Sigma(E)_\sigma=\sum_{k}\frac{V^2}{E-\varepsilon_k-<n^{-\sigma}>I+i\delta}$ is the spin-dependent self-energy~\cite{AP,half}. $G_{\sigma,a}$ is the Green's function of the adsorbate after it is adsorbed to the surface. We have ignored the band index and also assumed the V to be independent of k. The adsorption energy can further be written by using Eq.\ref{EB1} as,
\begin{equation}
\begin{split}
\Delta E (\delta m) & =-\frac{1}{\pi}\sum_{\sigma} \int^{E_F} E \Im \left[(1-\frac{d\Sigma(E)_\sigma}{dE})G_{\sigma,a}\right]dE+E_F\delta m\\
&+\frac{2E_F}{\pi}\int^{E_F}\Im\left[(1-\frac{d\Sigma(E)_\uparrow}{dE})G_{\uparrow,a}\right]dE 
+\sum_\sigma n_{a\sigma}(E_F-\varepsilon_{a\sigma})
\end{split}
\label{Almost-final}
\end{equation}
The self-energy, $\Sigma(E)_\sigma$ has real and imaginary parts and is usually given by $\Sigma(E)_\sigma=\Lambda (E)_\sigma-i\Delta(E)_\sigma$. Where $\Delta(E)_\sigma=\pi V^2 \sum_k \delta(E-\varepsilon_k-<n^{-\sigma}>I)=
\pi V^2 \sum_k \delta(E-\varepsilon_{k,\sigma})=\pi V^2 D_\sigma(E)$ is the imaginary part. The real part is obtained through the Hilbert transform: $\Lambda_\sigma(E)=\frac{1}{\pi} P \int \frac{\Delta_\sigma(E)}{E-E'}dE'$. Eq.\ref{Almost-final} gives the adsorption energy in terms of chemisorption induced surface magnetic moments. To estimate the mutual dependence of surface moments and adsorption energy, one needs to solve the Eq.\ref{mom2}, Eq.\ref{change} and Eq.\ref{Almost-final} simultaneously.  
\subsection{Numerical calculations and results}
\subsubsection{Chemisorption phase diagrams}
To understand the effect of magnetism on chemisorption and vice versa we consider a simple example: The chemisorption of an adsorbate with a single energy level, $\varepsilon_a$ relative to the metal surface with with occupation $n_a$. To obtain realistic results for our model calculation, we use realistic electronic structure as input for the calculation of the self energy $\Sigma(E)_\sigma$. We used the electronic structure of a bcc (110) film of Iron (Fe) to calculate the real and imaginary part of the self energies ($\Lambda (E)_\sigma$, $\Delta(E)_\sigma$). The Fe (110) surface was modelled as slabs of $2\times2$ in-plane unit cells and four atomic layers containing 16 atoms.
Studies of adsorption energies using four monolayers of metal were shown to be useful in some other studies~\cite{slab1,slab2}. We performed first-principles calculations are within the frame-work of Density Functional Theory (DFT) with Perdew-Burke Ernzerhof exchange correlation energy functional \cite{pbe} based on a generalized gradient approximation. We used a projector augmented wave method as implemented in Vienna ab-initio simulation package (VASP)~\cite{vasp}. Kohn-Sham wave functions of the valence electrons were expanded in plane wave basis with energy cut-off of 500 eV. The Brillouin zone sampling was carried out using Monkhorst Pack grid of $5\times 5\times 1$ k-points. Ionic relaxation was performed using conjugate-gradient method, until forces on unconstrained atoms were less than 0.04 eV/Angstrom for the non-constrained atoms. Vacuum of 10~$\AA$ was included. Dipole corrections were applied along the directions perpendicular to the metal surface in order to eliminate the unwanted electric fields arising from the asymmetry of the simulation cell. The structural relaxation were performed for only the top most two Fe layers. The bottom two layer are fixed to their bulk values. To obtain the inputs for the model calculation using Eq.\ref{change} and Eq.\ref{Almost-final}, the electronic structure inputs of the non-spin polarized calculations were used.
\par
\begin{figure}
\subfigure[]{\includegraphics[scale=0.65]{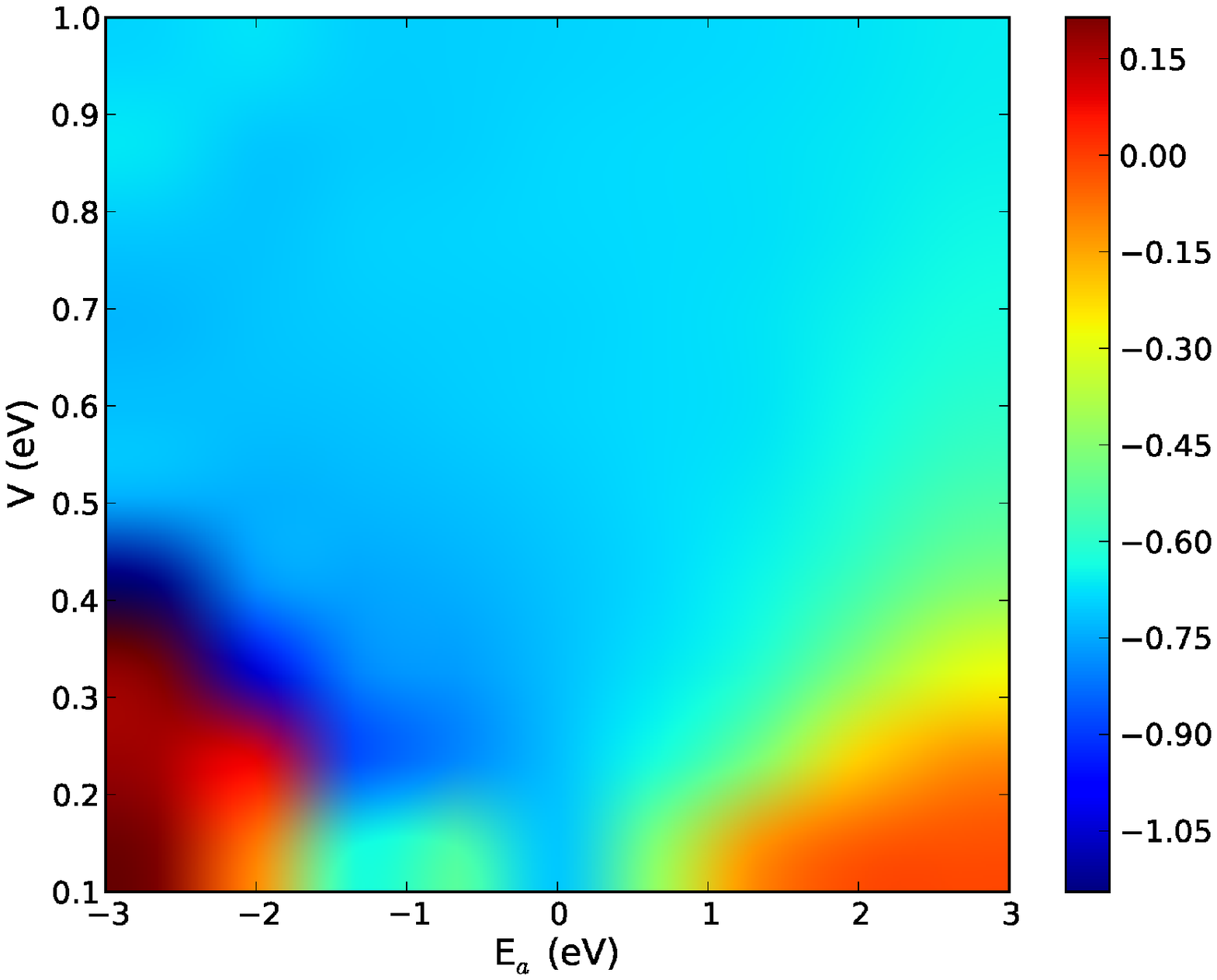}}
\subfigure[]{\includegraphics[scale=0.65]{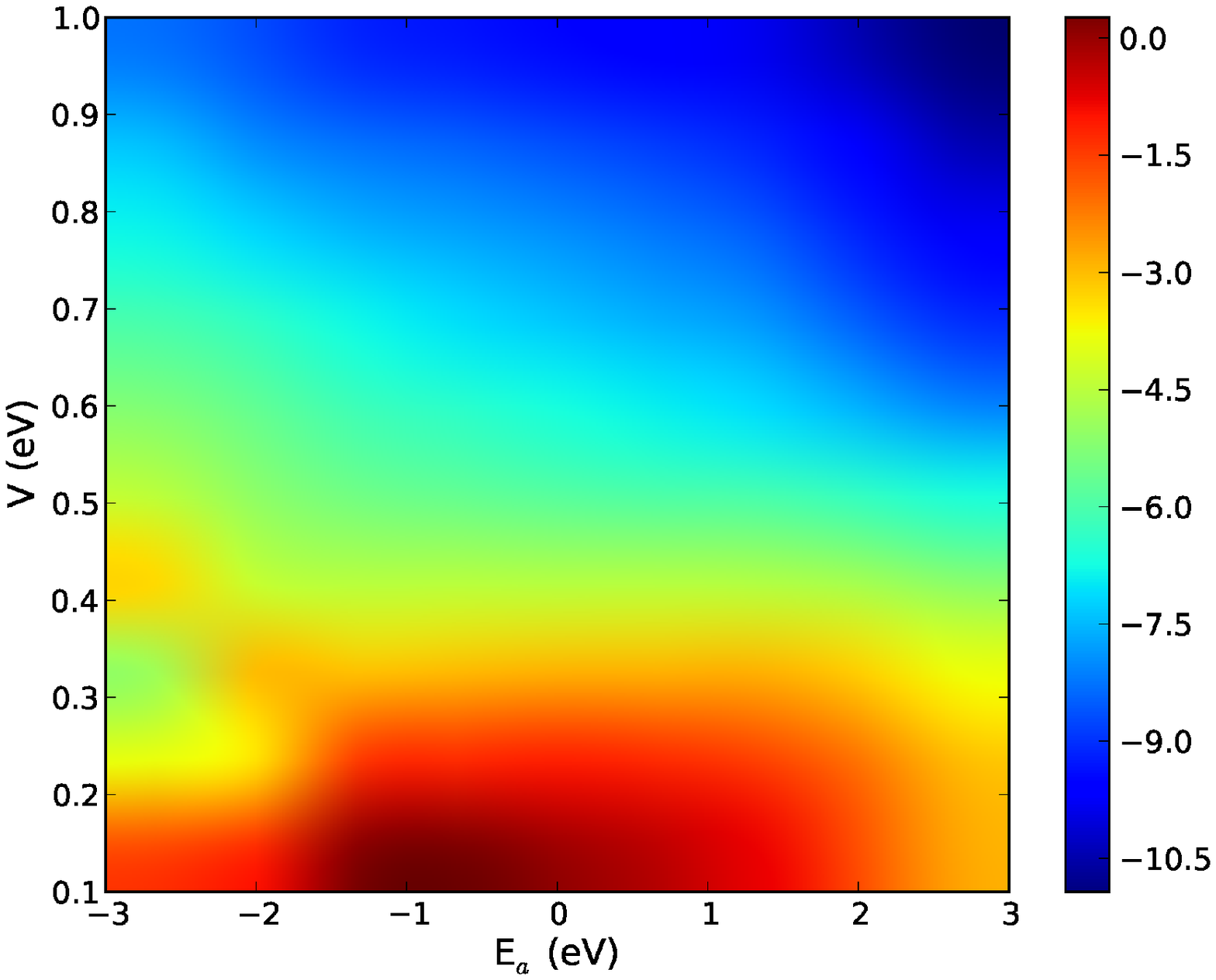}}
\caption{(a)Change in magnetic moment of the surface $\delta m$ (b) corresponding adsorption energies calculated from the Eq.\ref{Almost-final} for for different values of the coupling constant (V) and adsorbate energies ($\varepsilon_a$)}
\label{Fig1}
\end{figure}

In the Fig.\ref{Fig1}(a), we show the adsorbate induced change in magnetic moment ($\delta m$) which are calculated by simultaneously solving the  Eq.\ref{mom2} and Eq.\ref{change}. From Eq.\ref{change}, it can be understood that $\delta m$ depends on three parameters, the coupling constant V, the adsorbate energy level $\varepsilon_a$ and Stoner parameter I. The Stoner parameter was calculated using a fixed spin moment calculation (refer the supplemental material). In the Fig.\ref{Fig1}(a), we show such dependence in the form of a magnetic phase diagram. By sweeping different values of V and $\varepsilon_a$, we identify different regions where surface magnetic moments are reduced/enhanced due to the chemisorption. The initial magnetic moment per Fe-atom was set to 2.2 $mu_B$. We consider the adsorbate occupation as $\sum_\sigma n_{a\sigma}=1$. 
                    From the Fig.\ref{Fig1}(a), we can clearly notice the dependence of $\delta m$ on V and  $\varepsilon_a$. The general trend of the magnetic moment shows a decrement with increase in V . However for the values of V larger than 0.5 eV,the magnetic moment becomes almost intensive to V and varies very little. So it appears that effect of chemisorption on the surface magnetic moments has a critical value of the coupling, beyond which the moments are not effected much. $\delta m <0$ almost everywhere except $\varepsilon_a=-3 eV$ and $\varepsilon_a=-2 eV$ . It should be noted that most DFT-based adsorption studies of atomic adsorbates, such as hydrogen on ferromagnetic surfaces or ferromagnetic islands, report a decrease in magnetic moment~\cite{Jonsson}. However, one can see here that magnetic moment can also increase depending on the adsorbate's location in energy and the strength of the coupling. The quenching of the surface moment mainly correspond to the appearance of surface-adsorbate bonding peak near the Fermi energy, which increase the number of minority electrons.
\par
\begin{figure*}[]
%\subfigure[]{\includegraphics[scale=0.35]{3m.eps}}
%\subfigure[]{\includegraphics[scale=0.35]{3p.eps}}
%\subfigure[]{\includegraphics[scale=0.35]{0.eps}}
\includegraphics[scale=0.65]{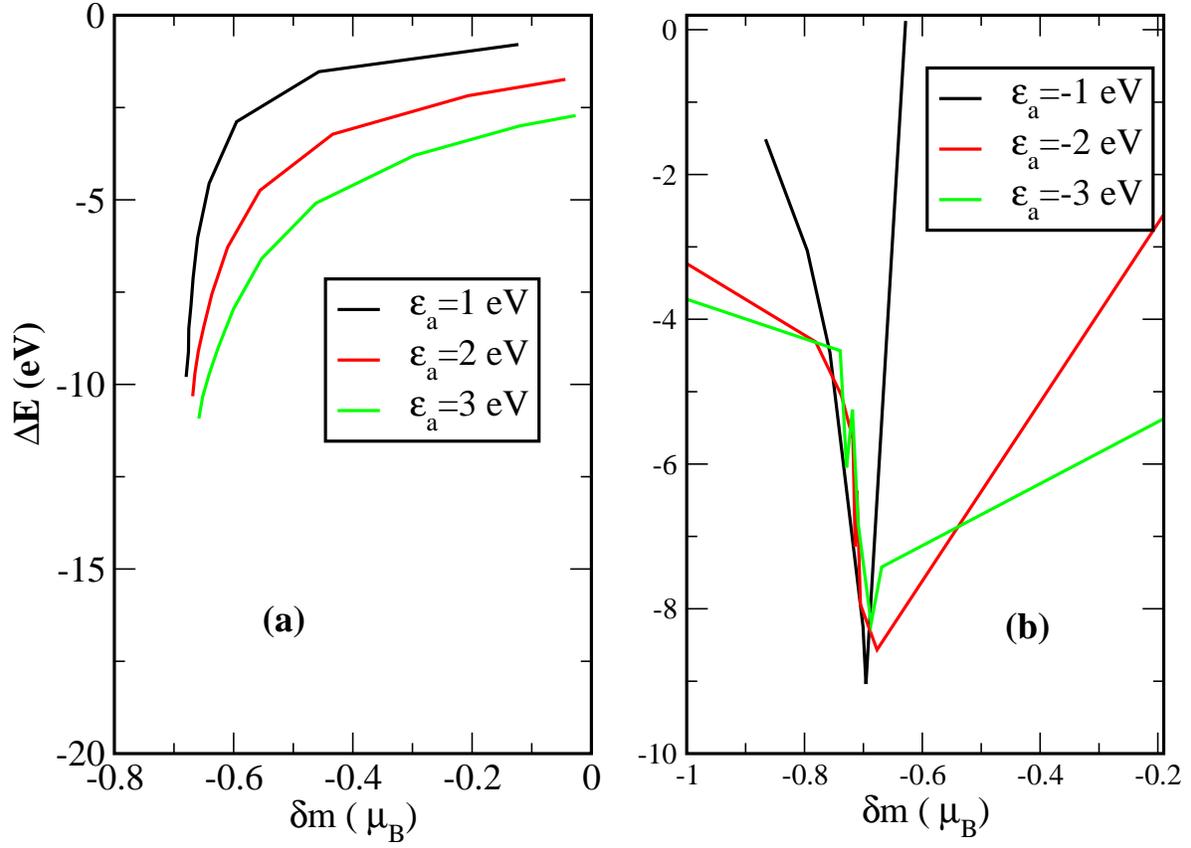}
\caption{(Color online) The change of adsorption energy with chemisorption induced magnetic moment $\delta m$. The results are shown for three values of $\varepsilon_a$. (a) for $\varepsilon_a> 0$ (b)$\varepsilon_a < 0$}
\label{Fig2}
\end{figure*}

In the Fig.\ref{Fig1}(b), we show the adsorption energy within the same regions  of V and $\varepsilon_a$ as in Fig.\ref{Fig1}(a). The adsorption energies are computed using Eq.\ref{Almost-final}.  We can see the adsorption energies follow the change of surface magnetic moments, however such the variation of adsorption energy with the change in magnetic moment are different for cases $\varepsilon_a > 0 eV$ to cases $\varepsilon_a < 0 eV$. 
For the cases when $\varepsilon_a > 0 eV$ the adsorption energy simply decrease as the magnetic moment increases while in the case of $\varepsilon_a < 0 eV$  when the magnetic moment increase, the adsorption energy first shows slight increment and finally increase.
Such behavior we further illustrate in the Fig.\ref{Fig2}, where we show the adsorption energy calculated  using Eq.\ref{Almost-final} with the induced moment $\delta m$ for $\varepsilon_a > 0 eV$ (Fig.\ref{Fig2}(a)) and $\varepsilon_a < 0 eV$ (Fig.\ref{Fig2}(b)). It can be noticed that in both cases the adsorption energy shows universal behaviour. In both cases $\Delta E$ can be written as a third order polynomial in terms of $\delta m$. This analysis therefore indicates some sort of scaling relationship of adsorption energies in terms of the surface magnetic moments, which we discuss below.
\subsubsection{Scaling relationships}
\begin{figure*}[]
\includegraphics[scale=0.65,keepaspectratio=true]{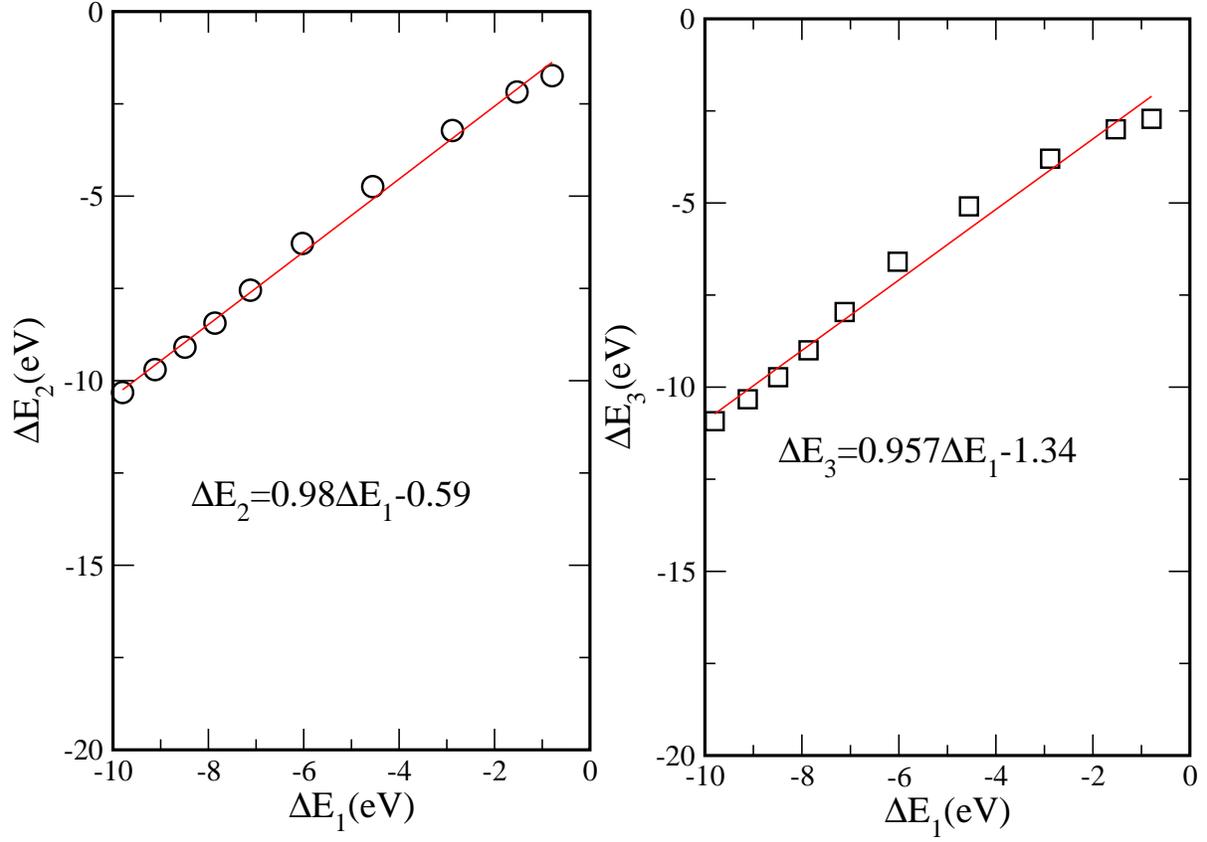}
\caption{(Color online) Linear scaling relationship between the adsorption energies corresponding the adsorbate with renormalized level $\varepsilon_a=1 eV$, $\varepsilon_a=2 eV$ and $\varepsilon_a=3 eV$ respectively.}
\label{Fig3}
\end{figure*}

The adsorption energies of different adsorbates within a family of similar adsorbates can be estimated by the use of so-called scaling relationships. Such scaling relationships minimize computational costs on the one hand, while on the other hand they place major restrictions on the design of the efficient catalysts. For an example, the adsorption energy, $\Delta E_{ads}^{AH_{x}}$ of an hydrogenated adsorbate AH$_x$ (A=C,N,O) can be related to the adsorption energy, $\Delta E_{ads}^{A}$ of the central atom A through~\cite{Greeley},
\begin{equation}
\Delta E_{ads}^{AH_{x}} = \gamma \Delta E_{ads}^{A} +\xi
\label{scaling}
\end{equation}
If we now imagine that the three levels with $\varepsilon_a < 0 eV$ as shown in Fig.\ref{Fig2}(a) correspond to three distinct adsorbates with distinct energy levels $\varepsilon_a=1 eV$, $\varepsilon_a=2 eV$ and $\varepsilon_a =3 eV$ respectively and see how their adsorption energies are related, we notice that they are linearly dependent to each other as can be seen from the Fig.\ref{Fig3}. The scaling relationship can be expressed as
\begin{equation}
\begin{aligned}
  \Delta E_2 &= \gamma_1\Delta E_1 +\xi_1 \\
  \Delta E_3 &= \gamma_2\Delta E_1 +\xi_2, 
\end{aligned}
\label{scale}
\end{equation}
The pool of surfaces that were considered here for the statistical analysis corresponds to the ones which are having induced magnetic moments ranging from -0.8$\mu_B$ to 0$\mu_B$ (refer Fig\ref{Fig2}(a)). Here $\Delta E_1$, $\Delta E_2$  and $\Delta E_3$ are respectively the adsorption energies for the adsorbate with renormalized energy levels 1,2 and 3eV respectively. The slopes $\gamma_1$ and $\gamma_2$ usually depend on the valencies of the adsorbates~\citep{scaling-o,scaling1}. However, as the valencies are same for all the three adsorbates here, according to the original formulation of the scaling relationship~\cite{scaling-o}, the slopes should be same for both cases (i,e $\gamma_1=\gamma_2$). But one can see that it is not the case here, as $\gamma_1=0.98$ and $\gamma_2=0.957$). This suggests the importance of the surface descriptors. As the pool of surfaces are considered here are based on the magnetic moments, the difference in $\gamma_1$ and $\gamma_2$ is expected to result from that. Recent studies suggest that the adsorption energies can be written in terms of a set of surface properties \cite{scaling2},
\begin{equation}
\Delta E_1 = F(\lbrace\omega_i\rbrace) + \alpha_{0} \quad
\Delta E_2 = G(\lbrace\omega_i\rbrace) + \beta_{0}
and \quad 
\Delta E_3 = H(\lbrace\omega_i\rbrace) + \gamma_{0}
\label{functions}
\end{equation}
where $F$, $G$  and $H$ are the functions of the set $\lbrace\omega_i\rbrace$ of certain surface properties and $\alpha_{0} $, $\beta_{0}$ $\gamma_0$ depend on surface coordination number. In the present case, one can obtain from the Fig.\ref{Fig2}(a) that for the case of $\varepsilon_a > 0$, the functions F, G and H can be obtained from $\Delta E_1=251\delta m+730\delta m^2+623.21\delta m^3+20.13$;  $\Delta E_2=22.5\delta m+95.3\delta m^2+121.65\delta m^3-0.82$; and $\Delta E_3=14.36\delta m+56.92\delta m^2+82.55\delta m^3-2.23$. As $\frac{F(\lbrace\omega_i\rbrace)}{G(\lbrace\omega_i\rbrace)}\neq \frac{F(\lbrace\omega_i\rbrace)}{H(\lbrace\omega_i\rbrace)}$, $\gamma_1\neq \gamma_2$. Therefore, it can be seen that the adsorption energy scaling depends on surface properties as well and for ferromagnetic surfaces the most potential descriptor could be the surface magnetic moment itself as is claimed by other recent study~\cite{Our}.\\
\par
The above discussions can be useful in terms of finding new route to enhance the catalytic activity in important electrochemical reactions such as oxygen evolution reaction (OER). The most important reaction intermediates for OER are *OH and *OOH which have the same valency. In most of the materials, the energy separation between *OH and *OOH is about 3.2 eV~\cite{Last}, and therefore according to the standard scaling relationship, the following relationship, $\Delta E_{OOH}=\Delta E_{OH}+3.2 eV$ holds. Now, for an ideal OER catalyst, the energy separation should be:$\Delta E_{OOH}-\Delta E_{OH}=2.46 eV$~\cite{Last}. It can be seen from the above by taking the statistics over a collection of magnetic surfaces where chemisorption induced change in magnetic moments only vary within the range of $0-0.8\mu_B$  brings a surface dependence  in slope. According to us, therefore, the \textit{surface independent} slope in linear scaling relationship is meaningful as long as the perturbation to the surface due to the chemisorption is negligible. 

\section{Chemisorption on non-magnetic surfaces}
\subsection{Modified Stoner Criterion due to the chemisorption}
Our next objective is to understand how typically a non-magnetic surfaces become ferromagnetic as is observed by \cite{beating,mujika}. For this, we consider a non-magnetic metal surface with density of states $D(E)$. Let us keep the notations similar to the above case of spin-polarized surface and just drop the spin index "$\sigma"$. As the surface is on-magnetic, it satisfies the usual Stoner condition $D(E_F)<1$, $E_F$ is the Fermi energy. If the surface becomes ferromagnetic when it  adsorbs a specific molecule, essentially the following condition has to be satisfied: 
$\left [D(E_F)+\Delta D(E_F)\right]I \geq 1$ that furthermore sets the condition in accordance with the Eq.(\ref{change}) as
\begin{equation}
\left [D(E_F)+D_a(E_F)+\frac{1}{\pi} \Im\lbrace \frac{d\Sigma(E)}{dE}G_{a}(E) \rbrace_{E_F}\right] I =1
\label{stoner-m1}
\end{equation}
%Now as $D(E_F)I <  1$, the modified Stoner criterion now reads,
%
%\begin{equation}
%\left (1-\left[\frac{d\Lambda}{dE}\right]_{E_F}\right)D_a(E_F) -\frac{1}{\pi} \left[\frac{d\Delta(E)}{dE}\right]_{E_F}Re\left[G_{a}\right] > 0
%\label{stoner-m2}
%\end{equation}
\subsection{Numerical results}
\begin{figure*}[]
\includegraphics[scale=0.65,keepaspectratio=true]{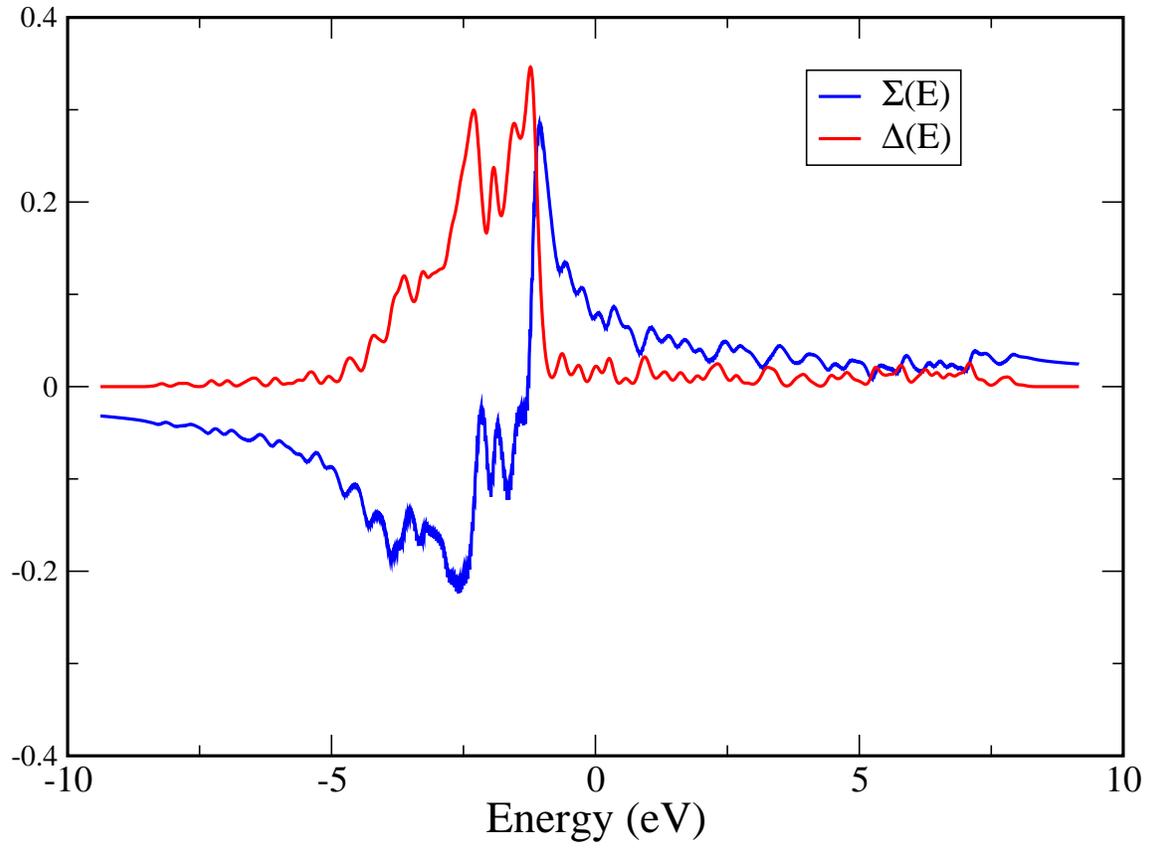}
\caption{(Color online) The real and imaginary of part of the self-energy obtained from the density of states of Cu (111) surface.}
\label{Fig4}
\end{figure*}

\begin{figure*}[]
\center
\includegraphics[scale=1.2]{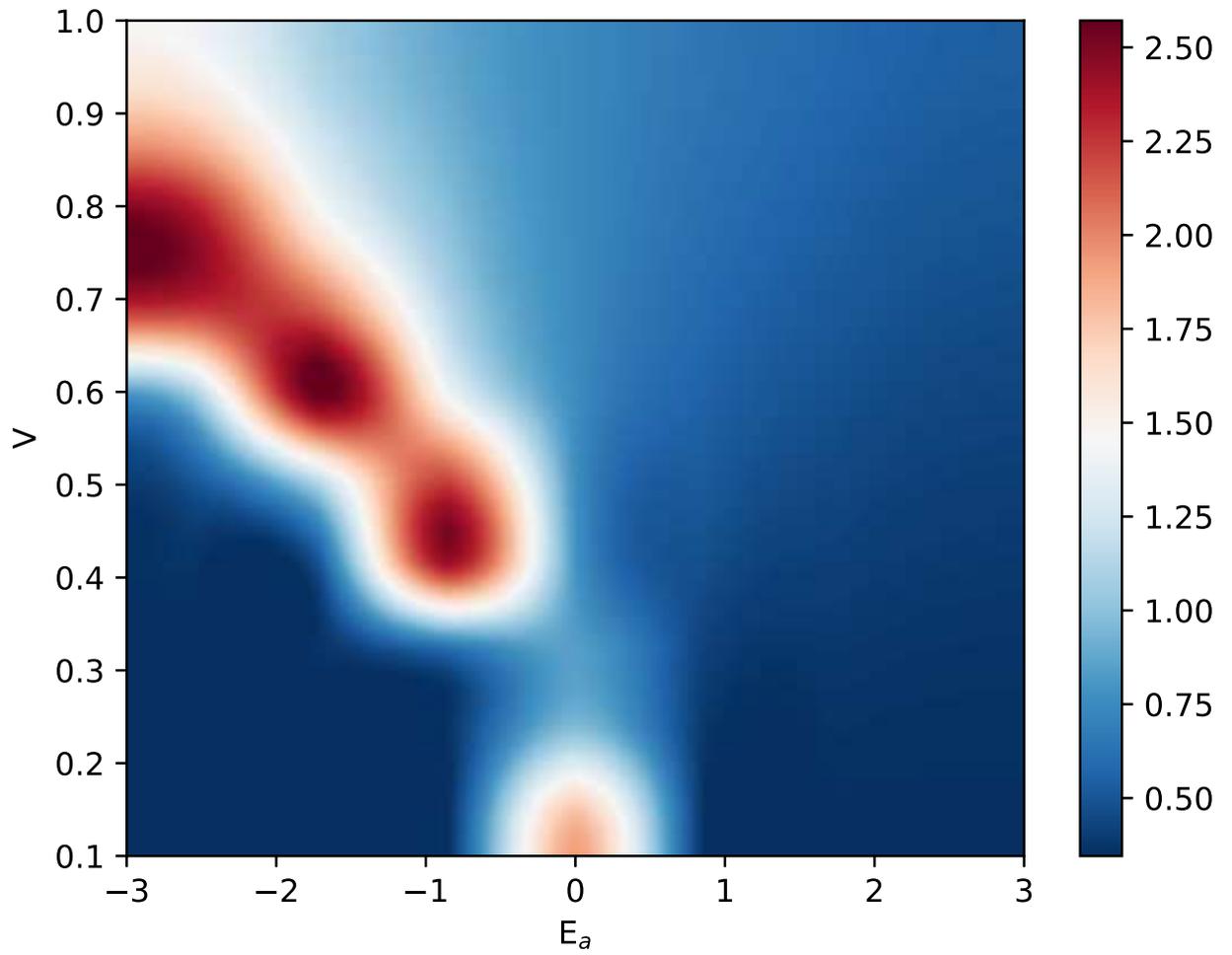}
\caption{(Color online) Heat-map of the left side of the Eq.\ref{stoner-m1} for different values of the coupling constant (V) and adsorbate energies (E$_a$). Red and yellow regions corresponds to ferromagnetism while the blue regions correspond to the non-magnetic state.}
\label{Fig5}
\end{figure*}

\begin{figure*}[]
\includegraphics[scale=0.7]{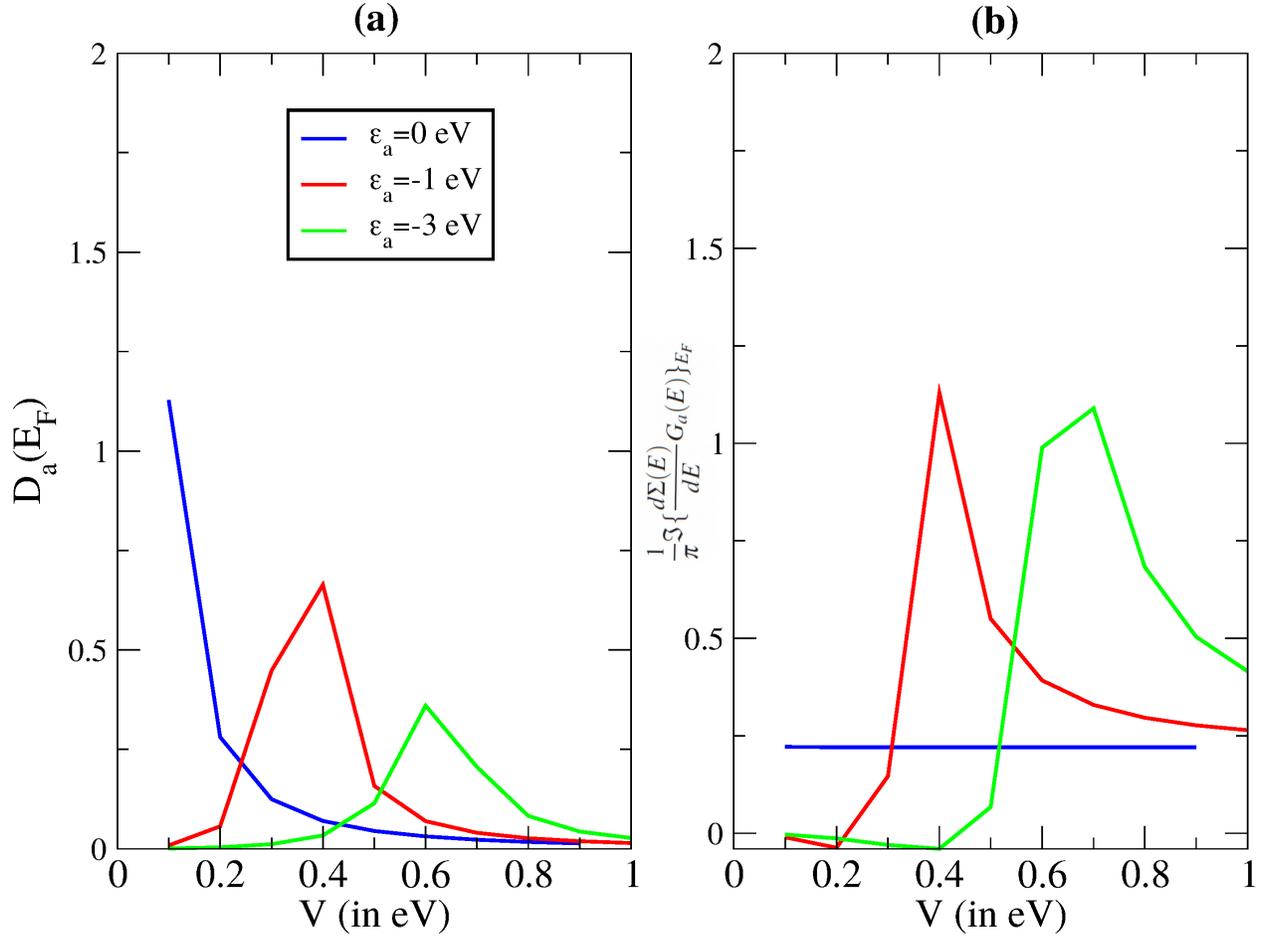}
\caption{(Color online) (a)The adsorbate DOS at Fermi energy (after chemisorption) as a function of V (b) (b)  $\frac{1}{\pi} \Im\lbrace \frac{d\Sigma(E)}{dE}G_{a}(E) \rbrace_{E_F}$ as a function of V. }
\label{Fig6}
\end{figure*}

The above equation can be thought of as a modified criterion for appearance of ferromagnetism 
in an otherwise non-magnetic surface via chemisorption. To demonstrate in a numerical way, we again consider an hypothetical atomic adsorbate whose energy level we vary from -3.0 eV to 3.0 eV and look at the behavior of the (modified) Stoner criterion. In order to be close to a realistic situation, we consider the electronic structure of (111) surface of Cu (copper) as input. The surface was modelled as slab of $2\times 2$ in-plane unit cells and four atomic layers of Cu. The bottom two layers are fixed to their bulk values. We used the DOS of such system as an input for the self energy $\Sigma(E)$. The density of states $D(E)$ appear in Eq.\ref{stoner-m1} corresponds to the the density of states of the top two layers of he Cu (111) slab. In the Fig.\ref{Fig4}, we show the real and imaginary part of the self energy, $\Sigma(E)$. We then compute the left side of the Eq.\ref{stoner-m1} for different values of the adsorbate energy and the coupling constants. The results are shown in the Fig.\ref{Fig5}. We used value of I=0.5 eV, for the Stoner parameter, which is close to the usual value for Cu~\cite{SP}. The blue regions correspond to the non-magnetic states (less than 1.0), while yellow and red regions are the ferromagnetic ones (greater than 1). 
\noindent First, we see that, for $\varepsilon_a > 0.5 eV$ the criterion is not satisfied for any value of V and the system remains non-magnetic. Next, we see that for each value of $\varepsilon_a < 0.5 eV$ the surface undergoes from non-magnetic to ferromagnetic transition for certain allowed values  of the coupling constant, V. The lower critical value of V, for a given $\varepsilon_a$ depends on $\varepsilon_a$ itself. Deeper the $\varepsilon_a$, higher the critical value is. The reason for such behavior lies on how  both $D_a(E_F)$ and $\frac{1}{\pi} \Im\lbrace \frac{d\Sigma(E)}{dE}G_{a}(E) \rbrace_{E_F}$ depends on V and $\varepsilon_a$. The Fermi energy is set at zero, here. It is seen that if the adsorbate level is close the Fermi energy of the metal and the coupling is very small, the modified Stoner criterion is satisfied. For very small coupling the metal-adsorbate bonding states are very close to the $E_F$ of the metal, and the Eq.\ref{stoner-m1} is satisfied due to the contribution from $D_a(E_F)$ (refer Fig.\ref{Fig6} (a)). For $\varepsilon_a <0 $ the modified Stoner criterion is mainly satisfied via $\frac{1}{\pi} \Im\lbrace \frac{d\Sigma(E)}{dE}G_{a}(E) \rbrace_{E_F}$ as can be seen from Fig.\ref{Fig6}(b). It can be seen that peak in $\frac{1}{\pi} \Im\lbrace \frac{d\Sigma(E)}{dE}G_{a}(E) \rbrace_{E_F}$ move to higher value of coupling as $\varepsilon_a$ goes deeper in energy.
\section{Conclusions}
In conclusion, we have studied the problem of chemisorption of molecules on  metal surfaces by using an approach that combines the Newns-Anderson-Grimely model with the Stoner model of metallic ferromagnetism. We studied for the ferromagnetic surfaces, how the strength of chemisorption is related to the magnitude of the surface moments and vice versa. We also discussed how chemisorption affects Stoner's criterion for the appearance of ferromagnetism and therefore allows the non-magnetic surface to become ferromagnetic. We have discussed how such a process depends on the position of the adsorbate level and the mixing strength of the adsorbate-metal levels. Even though, this study is more relevant to the adsorbates with single occupied levels, the physical insights that we gather should be useful for studying more complicated molecules on the metallic surfaces. 
%\section{Data availability}
%The data that support the findings of this work  are available within the article.

\section{Acknowledgment}
This work was supported by the Convergence Agenda Program (CAP) of the Korea Research Council of Fundamental Science and Technology (KRCF).             
%------------------------Figure here------------------------------------------------------------------
\newpage
\bibliography{citations} %You need to replace "rsc" on this line with the name of your .bib file
\bibliographystyle{unsrt} %the RSC's .bst file
\clearpage
\newpage
\clearpage
\newpage
\clearpage
\newpage
\clearpage
\newpage
\clearpage
\newpage
\clearpage
\newpage
\clearpage

\end{document}

% uncomment the following two lines for two-column version
%\twocolumn[\hsize\textwidth\columnwidth\hsize\csname
%@twocolumnfalse\endcsname

\title{Supplemental material of \\"Cooperation and competition between magnetism and chemisorption"}

\author{Satadeep Bhattacharjee and Seung-Cheol Lee}

\affiliation{Indo-Korea Science and Technology Center (IKST), Bangalore, India}
\date{\today}

%\pacs{}

\maketitle

\vspace{2mm}\def\b#1{\boldsymbol{#1}}

{\Large{Derivation of the Eq.3 of the manuscript}}\\

The adsorption energy is given by,
\begin{equation}
\begin{split}
\Delta E(\delta m) &=\sum_\sigma\lbrace \int^{E_F+\delta E_F}E\tilde D_\sigma(E)dE-\int^{E_F} E D_\sigma(E)dE-n_{a\sigma}\varepsilon_{a\sigma}\rbrace\\
&=\sum_\sigma\lbrace \int^{E_F} E\tilde D_\sigma(E)dE+\int_{E_F}^{E_F+\delta E_F}E\tilde D_\sigma(E)dE
-\int^{E_F} E D_\sigma(E)dE-n_{a\sigma}\varepsilon_{a\sigma}\rbrace\\ 
&=\sum_\sigma\lbrace \int^{E_F} E\Delta D_\sigma(E)dE+E_F\tilde D_\sigma(E_F)\delta E_F-n_{a\sigma}\varepsilon_{a\sigma}\rbrace\\ \nonumber
\end{split}
\end{equation}
Now using the charge neutrality condition,\\
$$\int^{E_F+\delta E_F}\tilde D_\sigma(E)dE-\int^{E_F} D_\sigma(E)dE=n_{a\sigma}$$\\
or 
$$\int^{E_F}\Delta D_\sigma(E)dE+\tilde D_\sigma(E_F)\delta E_F=n_{a\sigma}$$
we get,\\
\begin{equation}
\begin{split}
\Delta E(\delta m) &=\sum_\sigma \int^{E_F} E\Delta D_\sigma(E)dE-E_F\int^{E_F}\Delta D_\sigma(E)dE+\sum_\sigma E_F(n_{a\sigma}-\varepsilon_{a\sigma})\\ 
&=\sum_\sigma \int^{E_F} E\Delta D_\sigma(E)dE-E_F\int^{E_F}\Delta D_\uparrow(E)dE
-E_F\int^{E_F}\Delta D_\downarrow(E)dE\\ 
& +\sum_\sigma E_F(n_{a\sigma}-\varepsilon_{a\sigma})\\ \nonumber
\end{split}
\end{equation}
Now by adding and subtracting $E_F\int^{E_F}\Delta D_\uparrow (E)dE$ to right hand side of the above equation we get the Eq.3 of the manuscript.\\

{\Large{Calculation of the Stoner parameter for the Fe-(110) film}}\\

\begin{figure}
\includegraphics[scale=0.7]{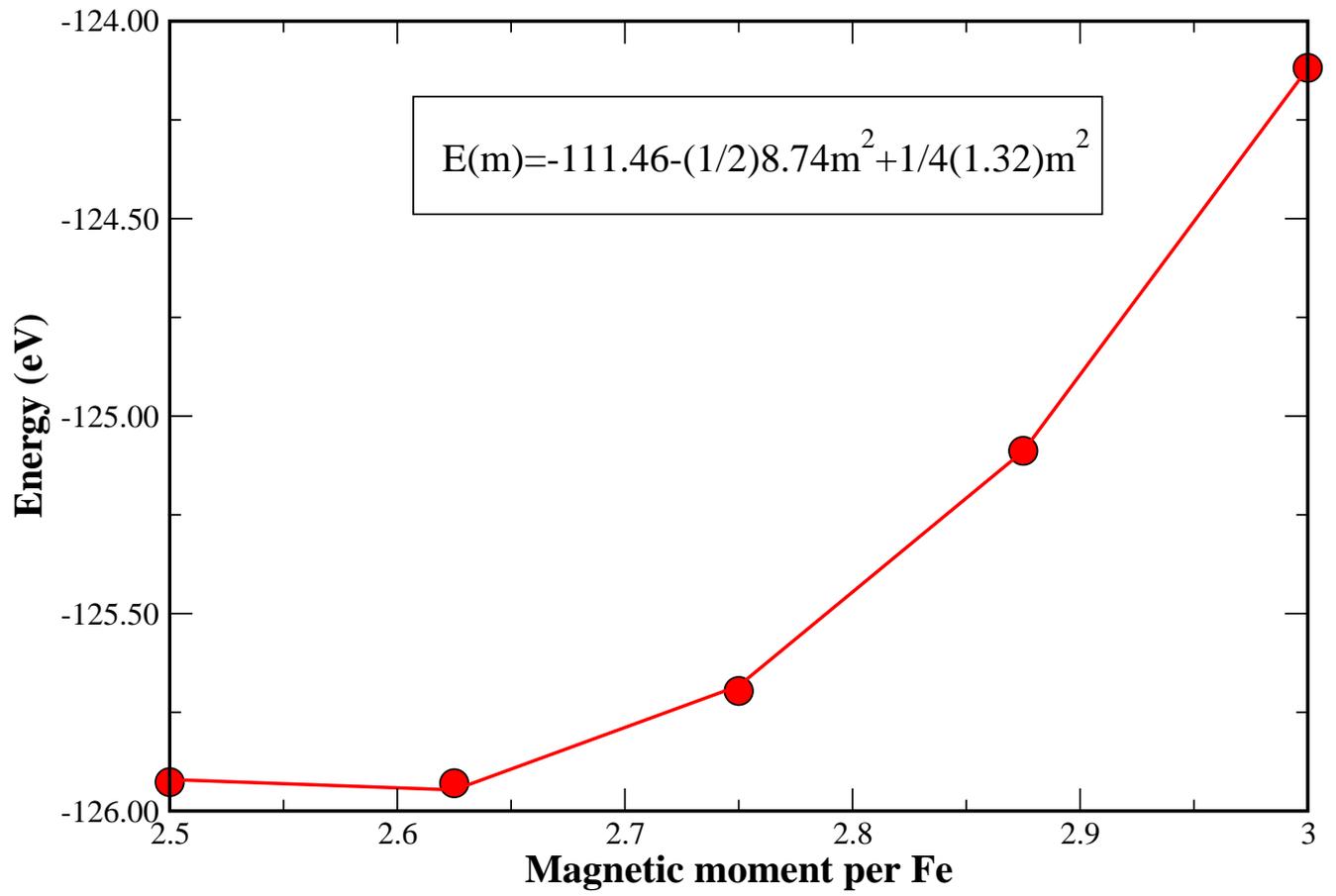}
\caption{(Color online) Fixed spin moment energy of bcc (110) surface as a function of the magnetic moment per Fe-atom}
\label{KK}
\end{figure}

We have have performed the fixed spin moment calculations for the Fe (110) film. The total energy
of the film can be written as,
\begin{equation}
E(m)=E(0)+\frac{1}{2} \chi^{-1}m^2+\frac{1}{4}\beta m^4
\label{LG}
\end{equation}
Where $E_0$ corresponds to the contributions from the non-magnetic degrees of freedoms, $\chi$ is the Stoner enhancement of the susceptibility. Therefore, $ \chi^{-1}=\frac{\partial^2E(m)}{dE^2}$. The Stoner parameter is obtained from the from the following~\cite{Ort}
\begin{equation}
I=\frac{1}{N(E_F)}-\frac{2}{\chi}
\end{equation}
We calculate $\chi$ by fitting the Eq.\ref{LG} to the total energy as shown in the Fig.\ref{KK}. The calculated value is 0.26.
\newpage
\clearpage
\newpage
{\Large{Derivation of the Eq.9 of the manuscript}}\\

The unperturbed Hamiltonian of the system is given by,
\begin{equation}
H_0=\sum_{k\sigma}\varepsilon_{k\sigma}c^\dagger_{k\sigma}c_{k\sigma}+\sum_\sigma \varepsilon_{a\sigma}n_{a\sigma}
\end{equation}
Where the first term represents the metal Hamiltonian, while the second term is that of the adsorbate. The band index is omitted. The Green's function of the the unperturbed system is given by,
\begin{equation}
G^0_\sigma=\frac{1}{E-H_0+i\delta}
\end{equation}
When we switch on the coupling between the two subsytems, the total Hamiltonian becomes,
\begin{equation}
H=H_0+V\sum_{k,j,\sigma}c^{\dagger}_{k\sigma}c_{a\sigma}+h.c
\end{equation}
The Green's function for the composite system is given by,
\begin{equation}
G_\sigma=\frac{1}{E-H+i\delta}
\end{equation}
For each spin, the change in density of states of the metal surface is given by,
\begin{equation}
\Delta D_\sigma(E)=\Delta n_\sigma (E)+\sum_\sigma \delta(\varepsilon-\varepsilon_{a\sigma})
\end{equation}
Where,
\begin{equation}
\begin{split}
\Delta n_\sigma(E) &=-\frac{1}{\pi}Im\left(G_\sigma(E)-G_\sigma^0(E)\right)\\
&=-\frac{1}{\pi}Im\left(\frac{d}{dE}ln det (1-V G_\sigma^0)\right)\\
&=-\frac{1}{\pi}Im \frac{d}{dE} ln(1-V^2\sum_k G_{k\sigma}^0G_{a\sigma}^0)
\end{split}
\label{long}
\end{equation}
Where
$$
G^0_{k\sigma}=\frac{1}{E-\varepsilon_{k\sigma}+i\delta}
$$
and 
$$
G^0_{a\sigma}=\frac{1}{E-\varepsilon_{a\sigma}+i\delta}
$$
are free metal and adsorbate Green's function respectively
Therefore from Eq.\ref{long}
\begin{equation}
\begin{split}
\Delta n_\sigma(E)&=-\frac{1}{\pi}Im \frac{d}{dE} ln(1-V^2\sum_k \frac{1}{E-\varepsilon_{k\sigma}+i\delta}\frac{1}{E-\varepsilon_{a\sigma}+i\delta})\\
&=-\frac{1}{\pi}Im \frac{d}{dE} ln(1-\frac{\Sigma_\sigma(E)}{E-\varepsilon_{a\sigma}+i\delta})\\
&=-\frac{1}{\pi}Im \frac{d}{dE}\left[ln(E-\varepsilon_{a\sigma}-\Sigma_\sigma(E)-ln(E-\varepsilon_{a\sigma}+i\delta)\right]\\
\end{split}
\label{long1}
\end{equation}
 Using the identity below,
$$
\frac{d}{dx}lnu(x)=\frac{u'(x)}{u(x)}
$$
we get 
$$
\Delta n_\sigma(E)=-\frac{1}{\pi}Im \left[\frac{(1-\frac{d\Sigma_\sigma(E)}{dE})} {E-\varepsilon_{a\sigma}-\Sigma_\sigma(E)}\right]-\delta(E-\varepsilon_{a\sigma})
$$
after rearranging we get,
\begin{equation}
\begin{split}
\Delta n_\sigma(E)+\delta(E-\varepsilon_{a\sigma})&=-\frac{1}{\pi}Im \left[\frac{(1-\frac{d\Sigma_\sigma(E)}{dE})} {E-\varepsilon_{a\sigma}-\Sigma_\sigma(E)}\right]\\
\Delta D_\sigma(E) &=-\frac{1}{\pi}Im \left[\frac{(1-\frac{d\Sigma_\sigma(E)}{dE})} {E-\varepsilon_{a\sigma}-\Sigma_\sigma(E)}\right]\\
& =-\frac{1}{\pi}Im \left[(1-\frac{d\Sigma_\sigma(E)}{dE})G_{a\sigma}(E)\right]\\
\end{split}
\end{equation}
The above equation gives the change in DOS of the adsorbent for a spin $\sigma$ due to chemisorption. This is used in the manuscript to obtain the most important physical parameters such as adsorption energy, change in surface magnetic moments etc.

%------------------------Figure here-------------------------------------------------------------------------